\documentclass[universe,article,accept,moreauthors,pdftex]{Definitions/mdpi} 
\definecolor{red}{rgb}{0.8,0,0}
\definecolor{violet}{rgb}{0.4,0,0.4}
\definecolor{green}{rgb}{0,0.5,0.0}
\definecolor{navy}{rgb}{0.0,0.0,0.6}
\definecolor{orange}{rgb}{0.8,0.2,0.0}

\usepackage[normalem]{ulem}  

\firstpage{1} 
\pubvolume{7}
\issuenum{10}
\articlenumber{382}
\pubyear{2021}
\copyrightyear{2021}
\externaleditor{{Academic Editor:  Nicolas Chamel} 
} 
\datereceived{4 September 2021} 
\dateaccepted{8 October 2021} 
\datepublished{15 October 2021} 
\hreflink{https://doi.org/10.3390/\linebreak universe7100382} 

\Title{Equation of State and Composition of
  Proto-Neutron Stars and Merger Remnants with Hyperons }

\TitleCitation{Equation of State and Composition of
  Proto-Neutron Stars and Merger Remnants with Hyperons}

\Author{Armen Sedrakian 
 $^{1,2,}$*\orcidB{} and Arus Harutyunyan  $^{3,4}$\orcidA{}}

\AuthorCitation{Sedrakian, A.; Harutyunyan, A.}

\address{$^{1}$ \quad  Frankfurt 
 Institute for Advanced Studies,
Ruth-Moufang-Stra\ss e, 1, 60438 Frankfurt am Main, Germany\\
$^{2}$ \quad Institute of Theoretical Physics, University of Wroc\l{}aw,
50-204 Wroc\l{}aw, Poland \\
  $^{3}$ \quad Byurakan Astrophysical Observatory, Byurakan 0213, Armenia; arus@bao.sci.am \\
  $^{4}$ \quad   Department of Physics, Yerevan State University, Yerevan 0025, Armenia}

\AuthorNames{Armen Sedrakian, Arus Harutyunyan}
\corres{Correspondence: sedrakian@fias.uni-frankfurt.de}

\abstract{ 
Finite-temperature equation of state (EoS) and the composition
  of dense nuclear and hypernuclear matter under conditions
  characteristic of neutron star binary merger remnants and
  supernovas are discussed. We consider both neutrino
  free-streaming and trapped regimes which are separated by a
  temperature of a few MeV. The formalism is based on covariant
  density functional (CDF) theory for the full baryon octet with
  density-dependent couplings, suitably adjusted in the hypernuclear
  sector. The softening of the EoS with the introduction of the
  hyperons is quantified under various conditions of lepton fractions
  and temperatures. We find that $\Lambda$, $\Xi^-$, and $\Xi^0$
  hyperons appear in the given order with a sharp density increase
  at zero temperature at the threshold being replaced by an extended
  increment over a wide density range at high temperatures. The
  $\Lambda$ hyperon survives in the deep subnuclear regime. The
  triplet of $\Sigma$s is suppressed in cold hypernuclear matter up
  to around seven times the nuclear saturation density, but appears in
  significant fractions at higher temperatures, $T\geq 20$ MeV, in both
  supernova and merger remnant matter.  We point out that
  a special isospin degeneracy point exists where the baryon abundances
  within each of the three isospin multiplets are equal to each other
  as a result of (approximate) isospin symmetry. At that point, the
  charge chemical potential of the system vanishes. We find that under
  the merger remnant conditions, the fractions of electron and $\mu$-on
  neutrinos are close and are about 1\%, whereas in the supernova
  case, we only find a significant fraction ($\sim$10\%) of electron
  neutrinos, given that in this case, the $\mu$-on lepton number is
  zero.
 }


\begin{document}
\section{Introduction}
\label{sec:introduction}

Several astrophysical scenarios lead to the formation of hot,
neutrino-rich compact objects which contain nuclear and 
hypernuclear matter at finite temperature. One such scenario arises in the
core-collapse  supernova and proto-neutron star context, where a hot
proto-neutron star is formed during the contraction of the supernova
progenitor and subsequent gravitational detachment of the remnant from
the expanding ejecta~\cite{Prakash1997,Pons_ApJ_1999,Janka_PhysRep_2007,Mezzacappa2015,Connor2018ApJ,Malfatti:2019tpg,Burrows2020MNRAS}. 
A related scenario arises in the case of stellar black-hole formation when the
progenitor mass is so large (typically tens of solar masses) that the
formation of a stable compact object is not possible and a black hole is
inevitably formed~\cite{Sumiyoshi2007,Fischer2009,OConnor_2011,Schneider2020}.
Finally, the~binary neutron star mergers offer yet another scenario
where finite temperature nuclear and hypernuclear matter play an
important role~\cite{Shibata_11,Faber2012:lrr,Rosswog2015,Baiotti:2019sew}. In~the “hot” stage of evolution of these objects the thermodynamics 
of the matter is characterized by several parameters, for~example, density,
temperature and lepton fraction. This is in contrast to the case of
cold (essentially zero-temperature) compact stars whose thermodynamics
is fully determined by a one-parameter EoS relating
pressure to energy density under approximate $\beta$-equilibrium. 
An important feature of the hot stages of evolution of compact stars is
the trapped neutrino component above the trapping temperature
$T_{\rm tr}\simeq 5$~MeV---a regime where the neutrino mean-free-path
is shorter than the size of the star~\cite{Alford2018b}. 
As is well known, neutrinos affect
significantly the composition of matter and are important for the
energy transport and dynamics of supernova and binary neutron star
mergers. 

After the first observation of a massive compact star in
2010~\cite{Demorest2010Nature} which was followed by further
observations of such objects~\cite{Cromartie2020NatAs,Fonseca2016ApJ}
the interest in the covariant density functional (CDF) theories of
superdense matter resurged because its parameters became
subject to astrophysical constraints in addition to the (low-density)
constraints coming from laboratory nuclear physics~(for reviews
see~\cite{Oertel2017,Sedrakian2021,Burgio2021}). CDF based
models tuned to the astrophysical constraints that account for the
finite temperature, neutrino component, and~strangeness in the form of
hyperons appeared in recent
years~\cite{Oertel_PRC_2012,Colucci2013,Oertel_EPJA_2016,Marques_PRC_2017,Dexheimer_2018,Fortin_PASA_2018,Weber2019,Stone_2019,Roark2019,Raduta:2020fdn,Stone2021,Alford2021}.

In this work, we study the EoS and composition of nuclear and
hypernuclear matter both in the neutrino free and neutrino-trapped
regimes within the CDF formalism. Our numerical
implementation is based on that of
Ref.~\cite{Colucci2013} but also includes the hidden strangeness $\sigma^*$ and $\phi$ mesons which account
for the interactions amongst hyperons. In~addition, instead of using
SU(3) symmetry arguments of Ref.~\cite{Colucci2013} in the scalar
sector, we adjust the parameters to the depths of hyperon potentials,
as already done in Refs.~\cite{Lijj2018b,Lijj2019,Li2020PhRvD} in the
case of  zero-temperature  EoS.  In~this work, we use, 
for the sake of
conciseness, a~single nucleonic CDF with parameters chosen according
DDME2 parameterization~\cite{Lalazissis2005}. A~similar nucleonic
DDME2-model-based finite temperature EoS has been presented in
Ref.~\cite{Raduta:2020fdn}, where the  couplings in the hyperonic
sector were taken from 
Ref.~\cite{Fortin2016} which differ from the ones adopted here.  
In this work, we do not address microscopic models of hypernuclear
matter which predict too low masses associated for hyperonic stars,
see Refs.~\cite{Sedrakian2007PrPNP,Burgio2021} 
for~reviews.

This work is organized as follows. Section~\ref{sec:EoS_npe+H} is
devoted to the formal aspects of EoS and the composition of matter at
finite temperatures.  The~CDF formalism is discussed in 
Section~\ref{sec:Hyper_DDME2} and the choice of the baryon--meson coupling constants is addressed in Section~\ref{sec:couplings}. 
The thermodynamic conditions of baryonic matter relevant to neutron 
star mergers and supernovas are discussed in Section~\ref{sec:thermodynamics}.
Our numerical results are given in Section~\ref{sec:numerical_results}. 
Section~\ref{sec:conclusions} provides a short summary. We use the natural 
(Gaussian) units with $\hbar= c=k_B=1$, and~the metric signature 
$g^{\mu\nu}={\rm diag}(1,-1,-1,-1)$.

\section{Relativistic Density Functional with Density-Dependent~Couplings}
\label{sec:EoS_npe+H}
\unskip

\subsection{Equation of~State}
\label{sec:Hyper_DDME2}

We start with a description of the formalism of CDF as applied to
hyperonic matter. In~this work, we adopt the DDME2
parameterization~\cite{Lalazissis2005} which is based on the version
of the theory that uses density-dependent coupling constants for
the meson-baryon interactions~\cite{Typelparticles2018}. 

The Lagrangian of the stellar matter  is given by 
\begin{equation}
  \label{eq:Lagrangian}
  {\cal L} = {\cal L}_b +  {\cal L}_m +  {\cal L}_\lambda
  +  {\cal L}_{em},
\end{equation}
where the baryon Lagrangian is given by 
\begin{eqnarray}
  \label{eq:lagrangian} 
  {\cal L}_b \, =\,  \sum_b\bar\psi_b\bigg[\gamma^\mu \left(i\partial_\mu
-g_{\omega b}\omega_\mu
  -g_{\phi b}\phi_\mu - \frac{1}{2} g_{\rho B} \boldsymbol{\tau} 
\cdot \boldsymbol{\rho}_\mu\right) 
- (m_b - g_{\sigma b}\sigma 
  - g_{\sigma^* b}\sigma^*)\bigg]\psi_b ,
\end{eqnarray}
where the $b$-sum is over the $J_B^P = \frac{1}{2}^+$ baryon octet;
$\psi_b$ are the Dirac fields of baryons with masses $m_b$;
$\sigma,\sigma^*,\omega_\mu,\phi_\mu$, and~$\boldsymbol{\rho}_\mu$ are the mesonic 
fields and $g_{mb}$ are the coupling constants that are density-dependent. 
The $\sigma^*$- and $\phi$-meson fields only couple to hyperons. 
The mesonic part of the Lagrangian is given by

\begin{eqnarray}
{\cal L}_m &=& \frac{1}{2}
\partial^\mu\sigma\partial_\mu\sigma-\frac{m_\sigma^2}{2} \sigma^2 -
\frac{1}{4}\omega^{\mu\nu}\omega_{\mu\nu} + \frac{m_\omega^2}{2}
               \omega^\mu\omega_\mu - \frac{1}{4}\boldsymbol{\rho}^{\mu\nu}\cdot
\boldsymbol{\rho}_{\mu\nu}
               + \frac{m_\rho^2}{2} \boldsymbol{\rho}^\mu\cdot \boldsymbol{\rho}_\mu \nonumber\\
               &+& \frac{1}{2}
\partial^\mu\sigma^*\partial_\mu\sigma^*-\frac{m_\sigma^{*2}}{2} \sigma^{*2} -
\frac{1}{4}\phi^{\mu\nu}\phi_{\mu\nu} + \frac{m_\phi^2}{2}
               \phi^\mu\phi_\mu,
\end{eqnarray}
where $m_{\sigma}$, $m_{\sigma^*}$, $m_{\omega}$, $m_{\phi}$ and 
$m_{\rho}$ are the meson masses and $\omega_{\mu\nu}$, $\phi_{\mu\nu}$ and 
$\boldsymbol{\rho}_{\mu\nu}$ stand for the field-strength tensors of vector mesons
\begin{eqnarray}             
\omega_{\mu \nu}  = \partial_{\mu}\omega_{\nu} - \partial_{\mu}\omega_{\nu} ,\qquad
\phi_{\mu \nu}  = \partial_{\mu}\phi_{\nu} - \partial_{\mu}\phi_{\nu} ,\qquad
\boldsymbol{\rho}_{\mu \nu}  = \partial_{\nu}
\boldsymbol{\rho}_{\mu} - \partial_{\mu}\boldsymbol{\rho}_{\nu}.
\end{eqnarray}

 The leptonic Lagrangian is given by
\begin{eqnarray}             
{\cal L}_\lambda\,=\, \sum_{\lambda}\bar\psi_\lambda(i\gamma^\mu\partial_\mu -
      m_\lambda)\psi_\lambda,
\end{eqnarray}
where $\psi_\lambda$ are leptonic fields and $m_\lambda$ are their
masses. The~lepton index $\lambda$ includes electrons and $\mu$-ons.
In hot stellar matter, one needs to include also the three flavors of
neutrinos whenever they are trapped. An~approximate estimate of the
temperature above which neutrinos are trapped is $T_{\rm tr} = 5$~MeV.
We will neglect henceforth the strong magnetic fields present in certain 
classes of compact stars and drop the gauge part ${\cal L}_{em}$ 
of the Lagrangian. For~the inclusion of these effects see
Refs.~\cite{Sinha2013,Thapa:2020ohp,Dexheimer:2021sxs}. 
We do not consider in
this work the non-strange $J=\frac{3}{2}$ members of the baryons decuplet---the $\Delta$-resonances~\cite{Drago_PRC_2014,Cai_PRC_2015,Zhu_PRC_2016,Kolomeitsev2017,Sahoo_PRC_2018,Lijj2018b,Ribes_2019}; for a review, see~\cite{Sedrakian2021}.

The partition function $\cal Z$ of the matter can be evaluated in the 
mean-field and infinite system approximations from which one finds the
pressure and energy density 
\begin{eqnarray}
  P  =  P_m+P_b+P_{\lambda},\qquad 
          {\cal E} =  {\cal E}_m+{\cal E}_b+{\cal E}_{\lambda},
\end{eqnarray}
with the contributions due to mesons and  baryons given by
\begin{eqnarray}
P_m &=& - \frac{m_\sigma^2}{2} \sigma^2 -\frac{m_\sigma^{*2}}{2} \sigma^{*2}
          + \frac{m_\omega^2}{2} \omega_0^2 +  \frac{m_\phi^2 }{2} \phi_0^2
          + \frac{m_\rho^2 }{2} \rho_{03}^2,\\
{\cal E}_m &=& \frac{m_\sigma^2}{2} \sigma^2 +\frac{m_\sigma^{*2}}{2} \sigma^{*2} +
                                \frac{m_\omega^2 }{2} \omega_0^2 
                              +  \frac{m_\phi^2 }{2} \phi_0^2
                              + \frac{m_\rho^2 }{2} \rho_{03}^2,\\
  P_b &=& \frac{1}{3}\sum_{b} \frac{2J_{b}+1}{2\pi^2}\int_0^{\infty}\!
  \frac{dk\ k^4 }{E_k^{b}}
        \left[f(E_k^{b}-\mu_{b}^*)+f(E_k^{b}+\mu_{b}^*)\right],\\
    {\cal E}_b &=& \sum_{b} \frac{2J_{b}+1}{2\pi^2}  
\int_0^{\infty}dk \ k^2 E_k^{b} \left[f(E_k^{b}-\mu_{b}^*)
                   +f(E_k^{b}+\mu_{b}^*)\right] ,
\end{eqnarray}
where $2J_b+1 $ is the spin degeneracy factor of the baryon octet. The~lepton contribution is given by
\begin{eqnarray}
  P_{\lambda} &=&
                  \frac{1}{3} \sum_{\lambda} \frac{2J_{\lambda}+1}{2\pi^2}\int_0^{\infty}\!  \frac{dk\ k^4 }{E_k^\lambda} \left[f(E_k^{\lambda}-\mu_\lambda)+f(E_k^{\lambda}+\mu_\lambda)\right],\\
  {\cal E}_{\lambda} &=&  \sum_{\lambda}                         \frac{2J_{\lambda}+1}{2\pi^2}  \int_0^{\infty}
  dk\ k^2E_k^\lambda\left[f(E_k^\lambda-\mu_\lambda)
+f(E_k^\lambda+\mu_\lambda)\right],
\end{eqnarray}
where $2J_{\lambda}+1 =2 $ for electrons
and $\mu$-ons and $1$ for neutrinos of all flavors.
The single particle energies of baryons and leptons
are given by $E_k^{b} = \sqrt{k^2+m^{*2}_{b}}$  and
$E_k^{\lambda}=\sqrt{k^2+m_\lambda^2}$, respectively, where the effective (Dirac) 
baryon masses in the mean-field approximation are given by
\begin{equation}
m_{b}^* = m_b - g_{\sigma b}\sigma - g_{\sigma^*b}\sigma^*.
\end{equation}
Next, $f(E) = [1+\exp(E/T)]^{-1}$ is the Fermi distribution function at
temperature $T$.  The~effective baryon chemical potentials are given by
\begin{eqnarray}
\label{eq:mu_eff}
&& \mu_{b}^* = \mu_{b}- g_{\omega b}\omega_{0} - g_{\phi b}\phi_{0} 
- g_{\rho b}  \rho_{03} I_{3b} - \Sigma^{r}, 
\end{eqnarray}
where $\mu_{b}$ is the chemical potential,  $I_{3b}$ is the third component 
of baryon isospin and the 
rearrangement self-energy $\Sigma^r$, which arises from
density-dependence of the coupling constants, is given by
\begin{eqnarray}
  \Sigma^r=\sum_b \left(
  \frac{\partial g_{\omega b}}{\partial n_b} \omega_0 n_b+
 \frac{\partial g_{\rho b}}{\partial n_b}  I_{3b} \rho_{03} n_b+
    \frac{\partial g_{\phi b}}{\partial n_b}  \phi_0 n_b-
  \frac{ \partial g_{\sigma b}}{\partial n_b} \sigma n_b^s
  -  \frac{ \partial g_{\sigma^* b}}{\partial n_b} \sigma^* n_b^s
  \right).
\end{eqnarray}

In the mean-field approximation the meson  expectation
values are given by
\begin{eqnarray}
      &&  m_{\sigma}^2\sigma = \sum_{b} g_{\sigma b}n_{b}^s,
 \quad m_{\sigma^{*}}^2 \sigma^{*} = \sum_{b} g_{\sigma^{*}   b} n_{b}^s,\\
  && m_{\omega}^2\omega_{0}= \sum_{b} g_{\omega b}n_{b} ,
  \quad m_{\phi}^2\phi_{0}= \sum_{b} g_{\phi b}n_{b} ,\\
&&  m_{\rho}^2\rho_{03}= \sum_{b}  I_{3b} g_{\rho b} n_{b},
\end{eqnarray}
where the meson fields now stand for their mean-field values; the
scalar number density is given by
$n_{b}^s= \langle\bar{\psi}_b \psi_b\rangle$, whereas the baryon
number density is given by
$n_{b}= \langle\bar{\psi}_b \gamma^0 \psi_b\rangle$. 
Explicitly, they are given by
\begin{eqnarray}
\label{eq:density_b}
&&n_{b} = \frac{2J_b+1}{2\pi^2}\int_0^\infty\! k^2dk
 ~\left[f(E^b_k-\mu_b^*)-f(E^b_k+\mu_b^*)\right],\\
\label{eq:density_s}
&&n_{b}^s = \frac{2J_b+1}{2\pi^2}\int_0^\infty\! k^2dk\,\frac{m^*_b}{E_k^b} 
\left[f(E^b_k-\mu_b^*)+f(E^b_k+\mu_b^*)\right].
\end{eqnarray}

\subsection{Choice of Coupling~Constants}
\label{sec:couplings}

The coupling constants are functions of baryon density, $n_B$. This
accounts for modifications of interactions by the medium at zero
temperature; the extrapolation to finite temperature neglects the
influence of temperature on the self-energies of baryons at 
beyond-mean-field level. 
The nucleon--meson couplings are given by 
\begin{equation}
 g_{iN}(n_B) = g_{iN}(n_{\rm sat})h_i(x),
\end{equation}
where $n_{\rm sat}$ is the saturation density, $x = n_B/n_{\rm sat}$ and 
\begin{eqnarray} \label{eq:h_functions}
  h_i(x) =\frac{a_i+b_i(x+d_i)^2}{a_i+c_i(x+d_i)^2},~i=\sigma,\omega,\quad
h_\rho(x) = e^{-a_\rho(x-1)}.
\end{eqnarray}
For completeness, we list the values of parameters in Table~\ref{tab:1}.  
\begin{specialtable}[H]
\caption{ 
The values of parameters of the DDME2 CDF.}

    \setlength{\tabcolsep}{4.2mm}
\begin{tabular}{ccccccc}
\toprule
Meson ({$i$}) & $m_i$ (MeV) & $a_{i}$ & $b_{i}$ & $c_{i}$ & $d_{i}$ & $g_{iN}$ \\
\midrule
$\sigma$ & 550.1238 & 1.3881 & 1.0943 & 1.7057 & 0.4421 & 10.5396 \\
$\omega$ & 783 & 1.3892 & 0.9240 & 1.4620 & 0.4775 & 13.0189 \\
$\rho$ & 763 & 0.5647 &---&---&---& 7.3672 \\
\bottomrule
\end{tabular}
\label{tab:1}
\end{specialtable}

\pagebreak
 The density-dependent functions
$h_{i}(x)$ are subject to constraints $h_{i}(1)=1$,
$h_{i}^{\prime\prime}(0)=0$ and
$h^{\prime\prime}_{\sigma}(1)=h^{\prime\prime}_{\omega }(1)$.

Fixing the hyperonic coupling constants involves two sources of
information: (a)~the couplings of hyperons to the vector mesons are
chosen according to the SU(6) spin-flavor symmetric
model~\cite{Swart1963}; (b)~their couplings to the scalar mesons are
chosen such as to reproduce their phenomenological potential depths at
the saturation density, which are determined from experiments.

We express the hyperonic couplings in terms of their ratios to the corresponding couplings 
of nucleons: $R_{i Y}=g_{i Y}/g_{i N} $ for $i=\{\sigma,\omega,\rho\}$ and $R_{\sigma^* Y}
=g_{\sigma^* Y}/g_{\sigma N} $, \mbox{$R_{\phi Y}=g_{\phi Y}/g_{\omega N} $.} 
For $\Lambda$-hyperons, we adopt  $R_{\sigma\Lambda} =
0.6106$~\cite{Lijj2018b}, which is close to the value determined in
Ref.~\cite{Dalen2014} through fits to the $\Lambda$-hypernuclei.
The~likely range of the potentials
for $\Sigma$ and $\Xi$ hyperons are
\begin{eqnarray}
&&  -10\le U_{\Sigma}(n_{\rm sat}) \le 30 ~{\rm MeV},\\
&& -24\le U_{\Xi}(n_{\rm sat}) \le 0~{\rm MeV}, 
\end{eqnarray}
where the value $U_{\Xi}(n_{\rm sat}) = -24$~MeV has been given in
~\cite{Friedman2021} and is much deeper than the one expected from
Lattice 2019 results~\cite{Inoue2019AIPC,Sasaki:2019qnh}.  The adopted
values of the coupling constants are taken from Ref.~\cite{Lijj2018b}
and are listed in Table~\ref{tab:2}.  Note that it is implicitly
assumed that the couplings of mesons to hyperons have the same density
dependence as for nucleons.  The hidden strangeness mesons have masses
$m_{\sigma^*} = 980$ and $m_{\phi} = 1019.45$~MeV, with the density
dependence of their couplings coinciding with those of the couplings
of the $\sigma$- and $\omega$-mesons, respectively.

\begin{specialtable}[H]
  \caption{
    The ratios of the couplings of hyperons to mesons. See text for~explanations.}
    \setlength{\tabcolsep}{7.16mm}
\begin{tabular}{cccccc}
\toprule
$Y\backslash R$  & $R_{\omega Y}$  & $R_{\phi Y}$ & $R_{\rho Y}$
 & $R_{\sigma Y}$  & $R_{\sigma^* Y}$\\
  \midrule
$\Lambda$ &   2/3 &  $-\sqrt{2}/3$    & 0 & 0.6106 & 0.4777 \\
$\Sigma $ &   2/3 &  $-\sqrt{2}/3$    & 2 & 0.4426 & 0.4777 \\
$\Xi$     &   1/3 &  $-2\sqrt{2}/3$   & 1 & 0.3024 & 0.9554\\
\bottomrule
\end{tabular}
\label{tab:2}
\end{specialtable}

\subsection{Thermodynamic Conditions in Supernovas and Merger~Remnants}
\label{sec:thermodynamics}

Next, we adopt our hypernuclear CDF to the stellar conditions,
specifically to the cases of supernovas and binary neutron star mergers. As~already mentioned, two regimes arise depending on the
ratio of the neutrino mean-free-path to the size of the system: the
neutrino free regime in the case of this ratio being much larger than
unity, and~the trapped neutrino  regime in the opposite case.  Trapped
 neutrinos are in thermal equilibrium and are characterized
by appropriate Fermi distribution functions at the matter temperature.
Numerical simulations provide the lepton fractions that we adopt
in our static (time-independent) description. We assume that the
lepton number is conserved in each family, which implies that the
neutrino oscillations are neglected.  The~$\tau$-leptons are
neglected because of their large mass.  For~supernova matter,  the predicted electron and muon lepton numbers  are \mbox{typically $Y_{L,e}\,
\equiv Y_e+Y_{\nu_e} = 0.4$ and $Y_{L,\mu}\,\equiv Y_\mu+Y_{\nu_\mu} = 0$
~\cite{Prakash1997,Malfatti:2019tpg,Weber2019},}
where we introduced partial lepton densities normalized
by the baryon density $Y_{e,\mu} = (n_{e,\mu}-n_{e^+,\mu^+})/n_B$, where $e^+$ refers
to the positron and $\mu^+$---to the anti-muon.  Note, however, that $Y_{e}$
may vary significantly along with a supernova profile in a time-dependent
manner. Furthermore, muonization in the matter can lead to a small (of the order $10^{-3}$)
fraction of $\mu$-ons~\cite{Bollig_PRL_2017,Guo:2020tgx}
which we neglect
here. In~the case of neutron star mergers, the~hot remnant emerges
from the material of initial cold neutron stars, and~the lepton
fractions $Y_{L,e} = Y_{L,\mu}= 0.1$ are assumed  for the remnant of a
merger. The~adopted values reflect (approximately) those of the
pre-merger cold neutron~stars.

The stellar matter is in weak equilibrium and is charge neutral. The~equilibrium with respect to the weak processes requires 
\begin{eqnarray}
\label{eq:c1}
  &&\mu_{\Lambda}=\mu_{\Sigma^0}=\mu_{\Xi^0}=\mu_n=\mu_B,\\
\label{eq:c2}
 &&  \mu_{\Sigma^-}=\mu_{\Xi^-}=\mu_B-\mu_Q,\\
\label{eq:c3}
  &&\mu_{\Sigma^+}=\mu_B+\mu_Q,
\end{eqnarray}
where $\mu_B$ and $\mu_Q=\mu_p-\mu_n$ are the baryon and charge
chemical potentials, $\mu_b$ with
$b \in \{ n,p,\Lambda, \Sigma^{0,\pm}, \Xi^{0,-}\}$ are the  thermodynamic
chemical potentials of the baryons. 
The charge neutrality
condition is given in terms of 
the partial densities of charged baryons as
\begin{eqnarray}
n_p+n_{\Sigma^+}-(n_{\Sigma^-}+n_{\Xi^-})=n_Q.
\end{eqnarray}
Introducing the partial charge density normalized by the baryonic
density $Y_Q = n_Q/n_B$, the~charge neutrality condition can be
written
\begin{equation}
Y_Q = Y_e + Y_{\mu}.
\end{equation}
The free streaming and trapped
neutrino regimes are characterized by
\begin{eqnarray}
\label{eq:mu_condition}
  \mu_e &=& \mu_\mu =-\mu_Q = \mu_n-\mu_p, \quad \textrm{(free streaming)}\\
  \mu_e& = & \mu_{L,e} - \mu_Q, \quad \mu_{\mu} =  \mu_{L,\mu} -
  \mu_Q,
  \quad \textrm{(trapped)}
\end{eqnarray}
where $\mu_{L,e/\mu}$ are the lepton chemical potentials which are
associated with the lepton number $Y_{L,e} = Y_{e} + Y_{\nu_e}$ and
$Y_{L,\mu} = Y_{\mu} + Y_{\nu_\mu}$, which are conserved separately.
Combining the weak-equilibrium and charge neutrality conditions we are
now in a position to compute the EoS of stellar matter both in the
trapped and free streaming neutrino regimes. Note that it is implicitly
assumed that the matter is under detailed balance with respect to Urca
processes; if this condition is violated, then an additional ``isospin
chemical potential'' 
arises~\cite{Alford2019a,Alford2021b}. Additionally, note that we do
not constrain particles to their Fermi surfaces and any corrections
associated with the finite temperature features of the Fermi
distribution function are included in our $\beta$-equilibratium
condition.

\section{Numerical~Results}
\label{sec:numerical_results}

Our numerical procedure involves a solution of self-consistent equations
for the meson fields and the scalar and baryon densities for fixed
values of temperature, density, and~lepton numbers $Y_{L,e}$ and
$Y_{L,\mu}$, which are chosen according to the physical conditions
characteristic for supernovas and merger remnants, as~specified in
Section~\ref{sec:thermodynamics}. In~this work, we concentrate on the
features of EoS and particle fractions (or abundances) in the matter under
various thermodynamic~conditions.

\begin{figure}[H] 
\includegraphics[width=0.9\linewidth]{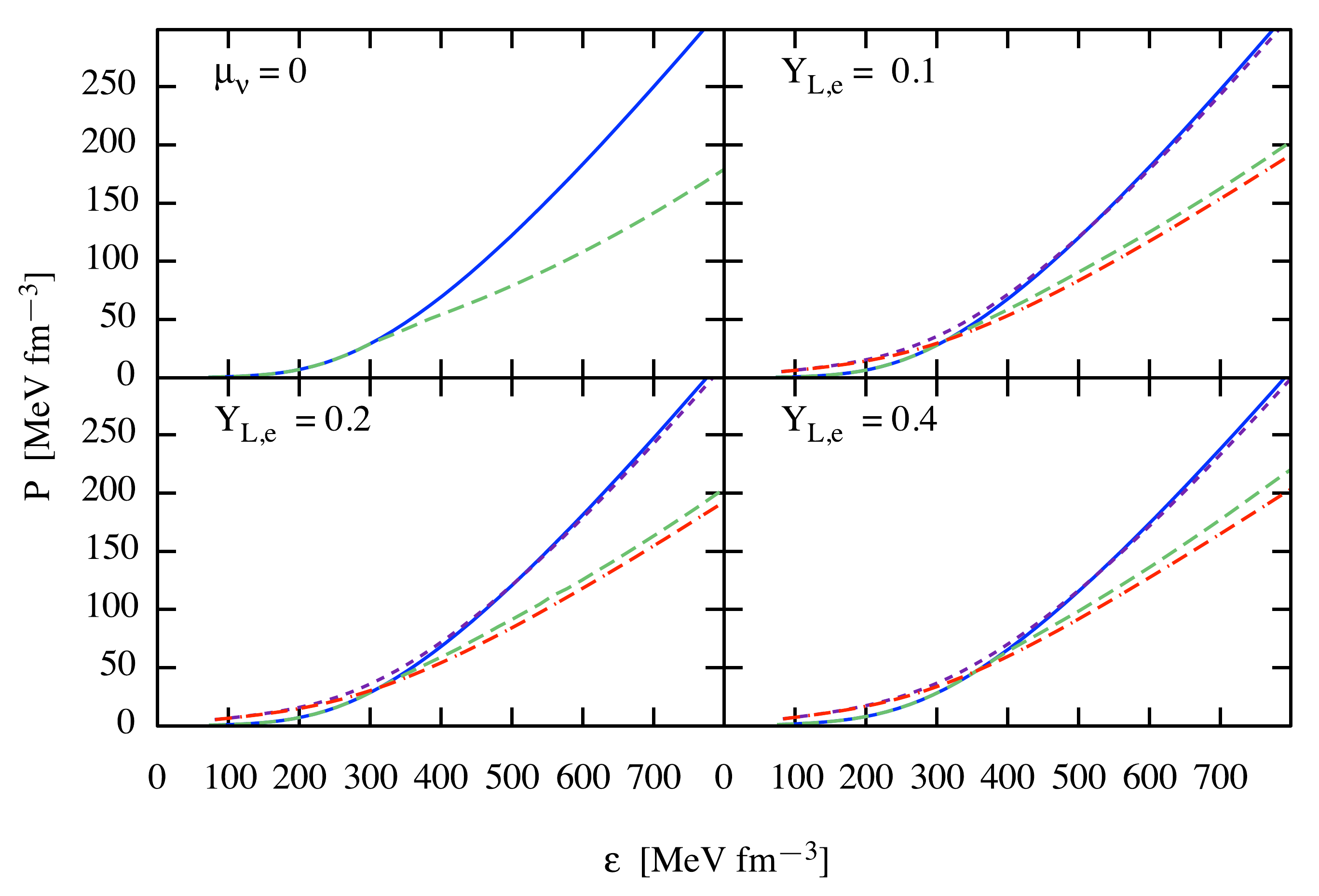}
\caption{Dependence of the pressure on the energy density.  The~panel
  labeled $\mu_\nu=0$ corresponds to neutrino-free $\beta$-equilibrium
  case without (solid) and with (dashed) hyperons at
  $T=0.1$~MeV. (Varying the temperature up to $T_{\rm tr}$ does not
  produce visible changes.)  The remaning panels show results for the
  neutrino trapped matter at $T=5$~MeV (solid---without hyperons and
  long-dashed---with hyperons) and 50~MeV (short-dashed---without
  hyperons and double-dash-dotted---with hyperons) for
  $Y_{L,e}= 0.1, \, 0.2, \, 0.4$ . The~$\mu$-on fractions are
  $Y_{L,\mu} = Y_{L,e} = 0.1$ (upper right panel) and $Y_{L,\mu} = 0$
  for $Y_{L,e}=0.2$ and 0.4 (lower row).  The~case $Y_{L,e} = 0.1$ is
  characteristic of a merger remnant, whereas
  $Y_{L,e} = 0.2,\, 0.4$---to supernova.}
\label{fig:EoS} 
\end{figure}

\textls[-15]{Figure~\ref{fig:EoS} shows the EoS for nucleonic and hyperonic matter
at \mbox{temperature $T=0.1$~MeV}} in the $\beta$-equilibrium and neutrino-free
case, as~well as at $T=5$ and 50~MeV with trapped neutrinos and
several values of $Y_{L,e}$.  The~$\mu$-on fractions are chosen as
$Y_{L,\mu} = 0$ for $Y_{L,e} = 0.2,\, 0.4$ and
$Y_{L,\mu} =Y_{L,e} = 0.1$. The~non-zero $Y_{L,\mu}$
is characteristic of merger remnants whereas zero $Y_{L,\mu}$ is characteristic for
supernovas. The~key well-known feature of the onset of hyperons seen
in Figure~\ref{fig:EoS} is the softening of the EoS, i.e.,~the shift of
pressure to lower values above the energy-density for the onset of
hyperons. It is further seen that for a higher temperature, the pressure
is larger at low densities and is lower at high densities independent
of the presence of~hyperons.

\begin{figure}[H] 
\includegraphics[width=0.9\linewidth]{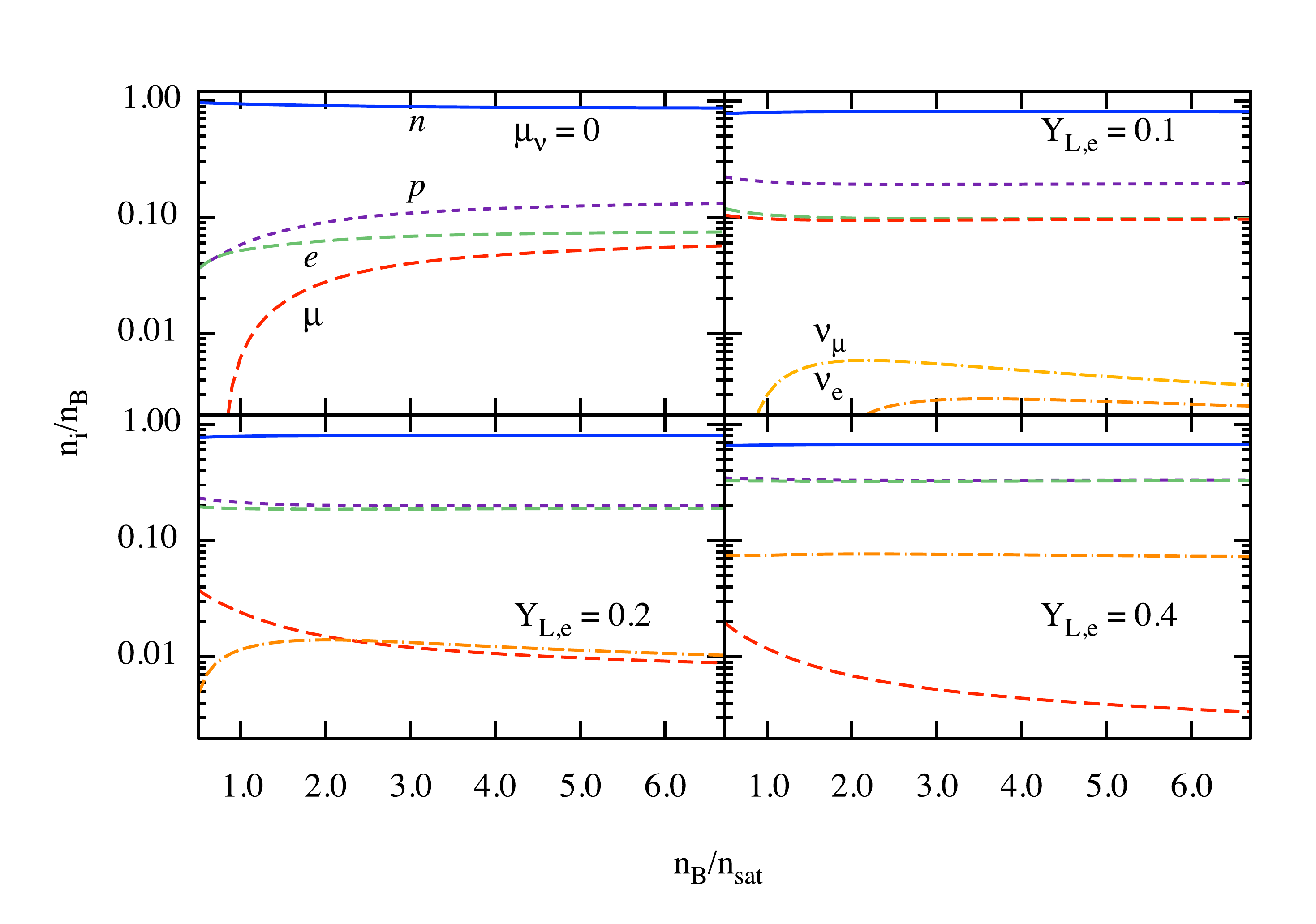}
\caption{Dependence of the particle fractions $n_i/n_B$ on the baryon
  density $n_B$ normalized by the saturation density. The~panels show
  the composition of $npe\mu$ matter in $\beta$-equilibrium at
  $T=1$~MeV in the neutrino-free case ($\mu_\nu = 0$) and for
  neutrino-trapped matter at $T=50$~MeV for several values of the
  electron lepton faction $Y_{Le} = 0.1,\, 0.2, \, 0.4.$, with~ the
  $\mu$-on component satisfying $Y_{L\mu} =Y_{Le} = 0.1$ and
  $Y_{L\mu} =0$ for $Y_{Le} = 0.2, \, 0.4$, where $Y_{L\mu}$ is the
  $\mu$-on lepton fraction.  }
\label{fig:abundances-npemu} 
\end{figure}

Figure~\ref{fig:abundances-npemu} shows the particle number densities
$n_i/n_B$ in $npe\mu$-matter normalized by baryon density as 
a function of baryon density normalized by
$n_{\rm sat} = 0.152$~fm$^{-3}$. The~case $\mu_\nu =0$ corresponds to
the  $\beta$-equilibrium neutrino-free case at $T=1$~MeV,
whereas the cases $Y_{L,e} = 0.1,\, 0.2, \, 0.4$ correspond to the
trapped neutrino regime at $T=50$~MeV.  The~choices of $Y_{L,\mu}$
match those of Figure~\ref{fig:EoS}.  In~contrast to the
neutrino-transparent case, where the muons appear above a threshold
density around $n_{\rm sat}$ where $\mu_e\geq m_\mu$, in~the
neutrino-trapped regime, the electron and muon contributions are almost
equal under merger conditions ($Y_{L,e} = 0.1$), and~there is a
visible fraction of $\mu$-on neutrinos.  Thus, the~charge neutrality
is maintained through the balance of negative charges of both types of
leptons with protons.  From~the upper right panel of
Figure~\ref{fig:abundances-npemu}, we see that the net neutrino numbers
become negative at low densities for both lepton families, indicating
that there are more antineutrinos than neutrinos in the low-density
and high-temperature regime of neutron star merger matter.

Note that the proton fraction remains below the
threshold for the Urca processes to operate in the low-temperature
neutrino-free regime. In~the high-temperature regime, the phase-space
for Urca processes opens due to the thermal smearing of Fermi
surfaces of baryons. This has important ramifications on the
oscillations of post-merger remnants through the damping effect of the
bulk viscosity driven by Urca processes~\cite{Alford2021,Alford2019a,Alford2019b,Alford2020a,Alford2021c}.
Under the supernova conditions, $\mu$-ons are greatly suppressed and the
corresponding neutrinos are extinct. Then, the~near equality of proton
and electron abundances is required by charge
neutrality. Note that the $\mu$-on abundances need not vanish,
as~$Y_{L,\mu}$ also includes the contributions from muonic neutrinos and
antineutrinos. The~small $\mu$-on fraction seen in the lower panels of
Figure~\ref{fig:abundances-npemu}
is compensated by an equal fraction of muonic antineutrinos $\bar{\nu}_\mu$
required by the condition $Y_{L,\mu}=0$.
The isospin asymmetry in supernova matter is reduced 
with increasing $Y_{L,e}$ and, consequently, the~difference between
the neutron and proton abundances gradually
vanishes. The~electron-neutrino population increases as well. 
In the cases $Y_{L,e} = 0.1 $, the $\mu$-on
neutrino fraction is comparable to that of electron-neutrinos, as
their lepton numbers are set equal. In~the lower panels of
Figure~\ref{fig:abundances-npemu}, they are absent because we
enforced the condition $Y_{L,\mu} = 0$.

\begin{figure}[H] 
\includegraphics[width=.91\linewidth]{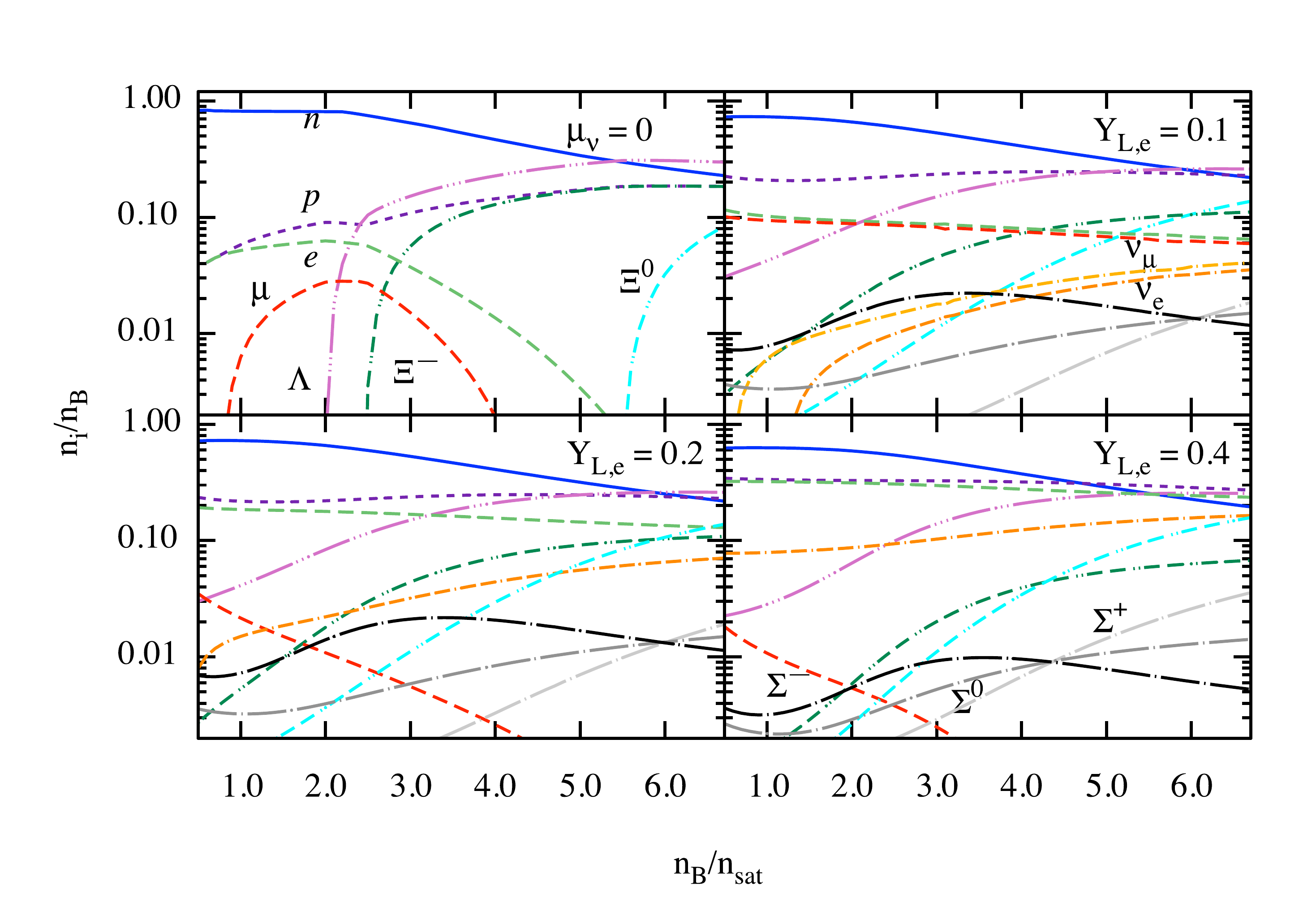}
\caption{ Same as in Figure~\ref{fig:abundances-npemu}, but~for the full
  baryon octet with the $\mu$-on component satisfying the conditions
  $Y_{L,\mu} =Y_{L,e} = 0.1$ (\textbf{upper right}) and $Y_{L,\mu} =0$ for
  $Y_{L,e} =0.2,\, 0.4$ (\textbf{lower row}). In~the low-temperature,
  $\beta$-equilibrium case, the $\Lambda$, $\Xi^-$ and $\Xi^0$ appear
  in the given order with a sharp increase in their fractions at  the corresponding density thresholds. 
  At high temperature $T=50$~MeV the density thresholds are located at lower
  densities (some are outside figure's scale) and the triplet 
  $\Sigma^{0\pm}$ appears. The~fractions of $\Lambda$ hyperons are shown by dash-triple-dot 
  lines, that of $\Xi^{0,-}$ by 
  double-dash-double-dot lines and that 
  of $\Sigma^{0,\pm}$ by dash-single-dot lines.  The~electron and $\mu$-ons
  neutrinos are shown by double-dash-dot lines; the electrons and $\mu$-ons 
  by long-dashed lines, protons by short-dashed lines and, finally, neutrons by solid lines. 
  }
\label{fig:abundances-hyperons} 
\end{figure}

Figure~\ref{fig:abundances-hyperons} shows the same as
Figure~\ref{fig:abundances-npemu}, but~it includes the full baryon
octet. Hyperons appear at densities above the saturation, in~the
following sequence: $\Lambda$, $\Xi^-$ and $\Xi^0$. The~onset of
$\Sigma^-$ hyperon in the low-temperature matter occurs at densities
outside the range shown.  The~reason for the shift of $\Sigma^-$
hyperons to high densities is the adopted highly repulsive potential
value in nuclear
matter~\cite{BartPhysRevLett,DOVER1984171,Maslov:2015wba,LopesPhysRevC2014,Gomes:2014aka,Miyatsu:2015kwa}. This
ordering is at variance to the case of free hyperonic gas, where
$\Sigma^-$ was predicted to be the first hyperon to
nucleate~\cite{Ambartsumyan1960SvA}, and more elaborate models which
assign weakly repulsive potential, see, e.g.,~\cite{Colucci2013}.
However, the~triplet of $\Sigma^{\pm,0}$ is present for $T=50$~MeV
independent of the values of lepton numbers.  It is interesting that
$\Sigma^-$ and $\Sigma^+$ fractions interchange their roles from being
most abundant to least abundant $\Sigma$-hyperon with increasing
density at a special intersection point where the abundances of all
the $\Sigma$s coincide. Note that the location of this special point
depends on the choice of $Y_{L,e}$. Furthermore, it is seen that the
intersection point of $n$ and $p$ fractions, as~well as that of
$\Xi^-$ and $\Xi^0$ fractions, are located close to the intersection point
of $\Sigma$s.

This feature can be understood by examining the
$\beta$-equilibrium conditions \eqref{eq:c1}--\eqref{eq:c3}.  If~there
is a point within the density range considered where the proton
fraction reaches the neutron fraction (which means $\mu_n^*=\mu_p^*$
due to Equation~\eqref{eq:density_b}), then the charge chemical potential
$\mu_Q=\mu_p-\mu_n$ vanishes at that point (due to the density
scaling~\eqref{eq:h_functions}, the contribution of the $\rho$-meson
mean-field to the effective baryon chemical
potentials~\eqref{eq:mu_eff} is negligible at high densities,
resulting in $\mu_n^*-\mu_p^*\simeq \mu_n-\mu_p$). This results in a
single chemical potential $\mu_b=\mu_B$ for the full baryon octet at
that special {\it isospin degeneracy} point. This implies, in~turn,
almost equal values of effective chemical potentials and, therefore,
{\it equal baryon fractions within a given~isospin-multiplet}.

\begin{figure}[H] 
\begin{center}
\hspace{-3cm}\includegraphics[width=.8\linewidth]{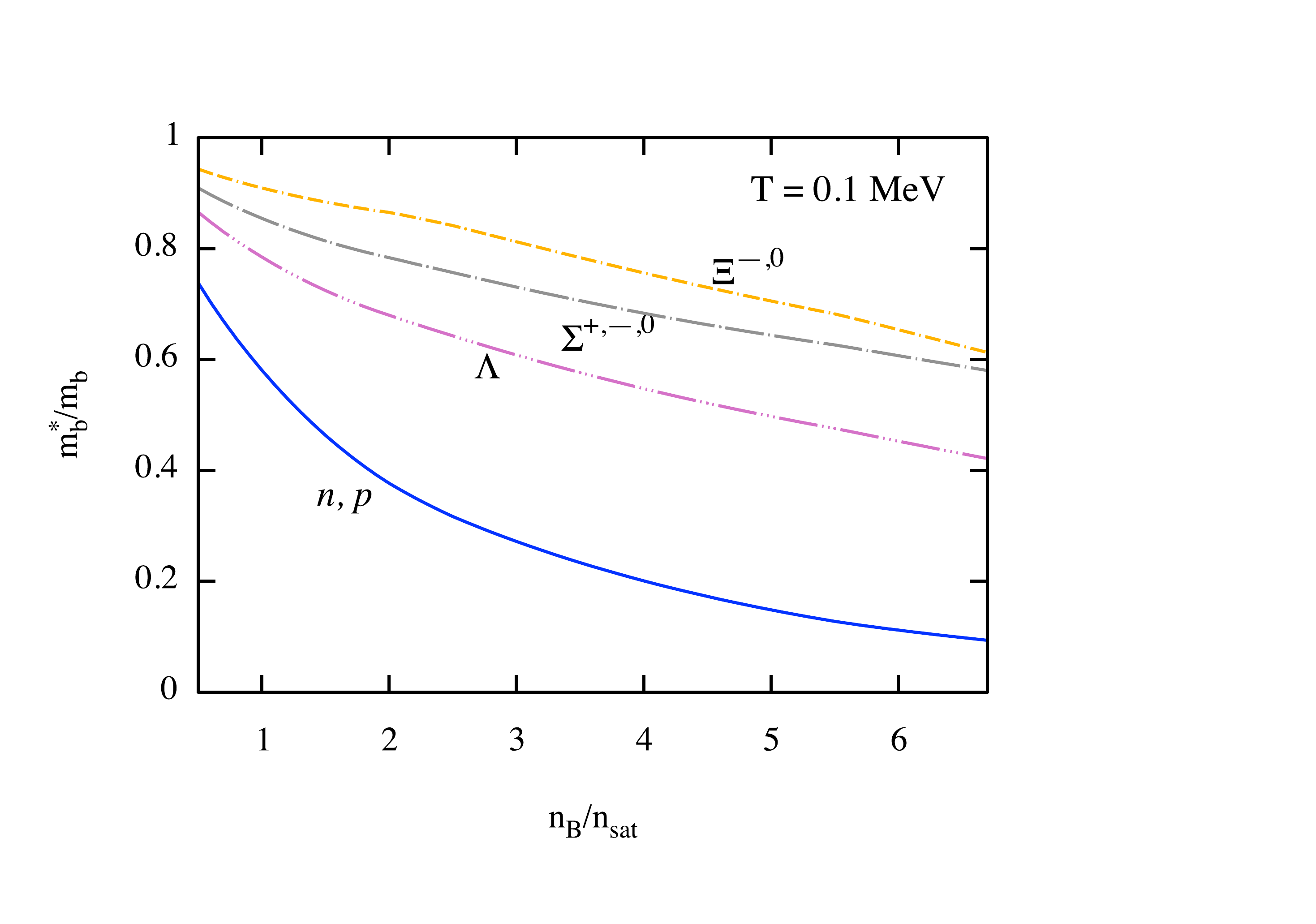}
\caption{ Dependence of effective masses of baryons on the density at $T=0.1$~MeV and in $\beta$-equilibrium. Each isospin multiplet is shown by a single line due to the degeneracy in their masses. 
}
\label{fig:meff-hyperons} 
\end{center}
\end{figure}

Figure~\ref{fig:meff-hyperons} shows the effective masses of baryons
as a function of density at $T=0.1$~MeV and in
$\beta$-equilibrium. The~effective masses of isospin multiplets
($n,p$), $\Sigma^{0,\pm}$ and $\Xi^{0,-}$
are degenerate. The temperature dependence of the effective masses of baryons is very
weak and, for~the sake of clarity, is not~shown.

\begin{figure}[H] 
\includegraphics[width=.9\linewidth]{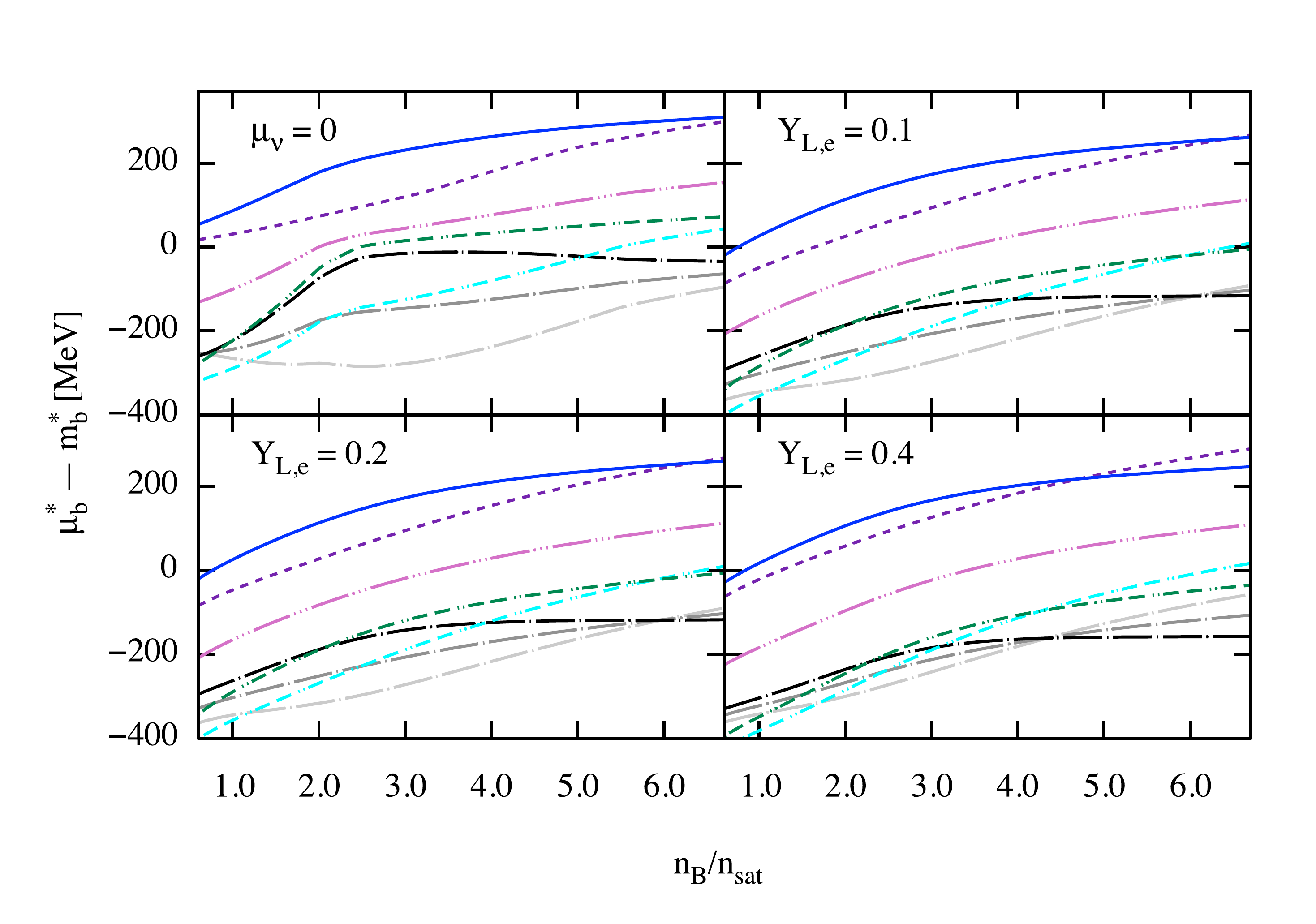}
\caption{ Dependence of baryon effective chemical potentials (computed
  from their effective masses) on the normalized baryon density
  $n_B/n_{\rm sat}$. The line styles for each baryon and the values of
  the temperature and lepton fractions for each panel match those in
  Figure~\ref{fig:abundances-hyperons}. The intersection (isospin degeneracy) points of
  chemical potentials of the same isospin-multiples is clearly visible
  at all values of lepton fractions in the neutrino-trapped matter.  }
\label{fig:mu-hyperons} 
\end{figure}

Figure~\ref{fig:mu-hyperons} shows the effective baryon chemical
potentials minus their effective masses, which clearly show the special
intersection points within each multiplet at all values of the lepton
fractions in the neutrino-trapped matter. Note that the effective masses
within each multiplet are equal in our model, see
Figure~\ref{fig:meff-hyperons}
above.  On~the left side of the
intersection point we have $\mu_Q\leq 0$, which according to the
conditions~\eqref{eq:c1}--\eqref{eq:c3} puts the baryon abundances
within each multiplet in the charge-decreasing order (i.e., baryons
with smaller charges are more abundant). Above~the intersection point
$\mu_Q\geq 0$, the ordering of baryon fractions within each
multiplet is reversed. Similar behaviour of baryon abundances was
found also in Refs.~\cite{Malfatti:2019tpg,Raduta:2020fdn},
where the composition of hot stellar matter was shown at constant
entropy-per-baryon and the composition of matter also included the
quartet of $\Delta$-resonances.  Note that in the ideal case of exact
isospin symmetry, the intersection points of the three
isospin-multiplets $n-p$, $\Sigma^{0,\pm}$ and $\Xi^{0,-}$ would be
located exactly at the same density. The~small deviations of these
three points from each other (which increase gradually with increasing
$Y_{L,e}$) reflect the fact that the isospin symmetry is approximate.

\vspace{-15pt}

The difference between almost equal abundances of leptons for
$Y_{L,e} = 0.1$ and the remaining cases $Y_{L,e} = 0.2,\, 0.4$ is
related to our choice of $Y_{L,\mu} $ to reflect merger remnant
conditions (first case) and supernova conditions (second case). This
difference also propagates to the abundances of electron and $\mu$-on
neutrinos, which are present in almost equal quantities in the first
case, whereas in the second case, the $\mu$-on neutrinos are replaced
by a much smaller amount of $\mu$-on antineutrinos. Hyperons affect the
way the charge neutrality is maintained at high density. In~low-temperature and $\beta$-equilibrated matter it is enforced by
equal abundances of protons and $\Xi^-$ hyperons with electrons and
$\mu$-ons being extinct at high densities.  At~finite temperature, the
electrons are abundant and the presence of $\Xi^-$ hyperon only induces some splitting between the electron and proton fractions, which
becomes less pronounced with increasing $Y_{L,e}$. The~fractions of
$\mu$-ons and their neutrinos in the merger remnant case
($Y_{L,e} = Y_{L,\mu} = 0.1$) are as significant as those of electrons
and electron-neutrinos, respectively, but~they do not play any
significant role in the supernova case where $Y_{L,\mu} = 0$.  In~contrast to the pure nucleonic matter where the neutrino abundances
remain constant or decrease slowly with baryon density, the~hypernuclear matter features increasing neutrino abundances with
density because of decreasing lepton fractions at fixed $Y_{L,e}$ and
$Y_{L,\mu}$.

It is further seen
that finite temperatures induce a significant shift of the hyperon
thresholds to lower densities (which lie outside of the density range
considered). This is in accordance with the recent observation that low-density hot nuclear matter may feature a significant fraction of
strangeness ($\Lambda$-particles) as well as $\Delta$-resonances in
addition to light clusters and free nucleons~\cite{Sedrakian:2020cjt}. 
Note also that the $\Lambda$-hyperon abundances become 
larger than 
those of neutrons at high density, i.e.,~these species are the 
dominant baryonic component in the matter for $n_B/n_{sat} \gtrsim 5.5$.
This results mainly from the~weaker repulsive coupling of $\Lambda$s to
$\omega$-meson which enhances their abundances compared to neutrons.
The weaker renormalization of $\Lambda$'s mass due to coupling to
$\sigma$ and $\sigma^*$ mesons than that of neutron is less important.

\begin{figure}[H] 
\includegraphics[width=.88\linewidth]{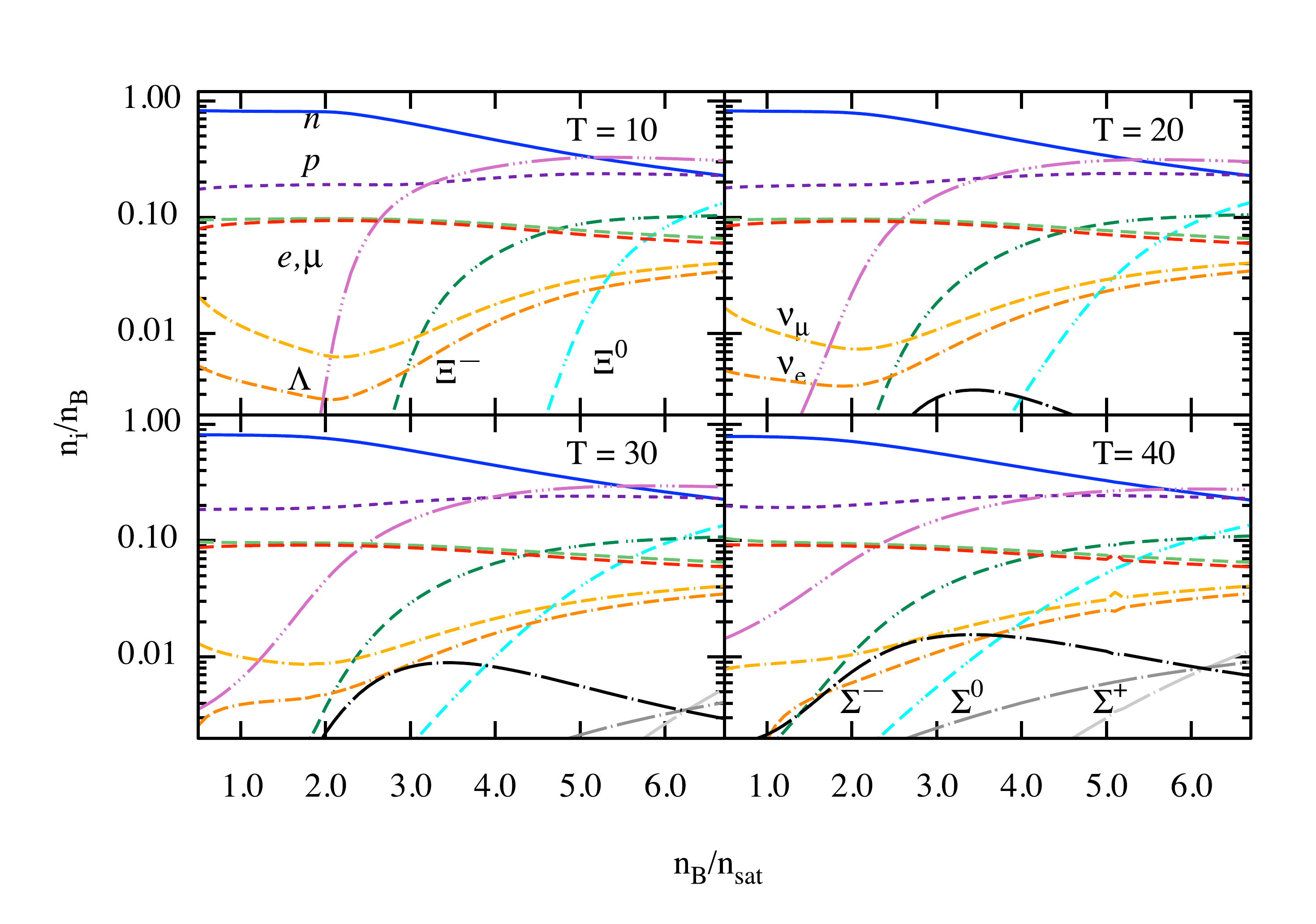}
\caption{ Same as in Figure~\ref{fig:abundances-hyperons}, but~for the
  fixed electron and $\mu$-on lepton fractions $Y_{L,e}= Y_{L,\mu}=0.1$ 
  and temperatures $T=10$, 20, 30 and 40 MeV. The~lepton number fractions 
  are characteristic for binary neutron star mergers. 
}
\label{fig:abundances-hyperons-T_dep_Y_01} 
\end{figure}

Figure~\ref{fig:abundances-hyperons-T_dep_Y_01} shows the particle
fractions in the hypernuclear matter in the temperature range
$ 10\le T\le 40$~MeV and electron and $\mu$-on fraction fixed by the
condition \mbox{$Y_{L,e}= Y_{L,\mu}=0.1$} characteristic of neutron star
binary mergers. It is seen that the abundances of neutrons, protons,
electrons and $\mu$-ons are weakly dependent on the
temperature. Due to equal lepton numbers, the electron and $\mu$-on
abundances are close to each other with the small electron excess
reflected in the dominance of $\mu$-on neutrinos over the
electron-neutrinos. At~high densities, the neutrino abundances 
are almost independent of the temperature as well, but~they decrease 
with increasing temperature and, eventually, become negative at 
temperatures between 40 and 50 MeV in the low-density domain 
(see also the upper right panel of Figure~\ref{fig:abundances-hyperons}).

Hyperons have sharply increasing fractions at the thresholds at
$T=10$~MeV, which replicates those at low temperatures and in neutrino-free
regimes. With~increasing temperature, the thresholds of the appearance
of the hyperons move to the lower densities, with the $\Lambda$ threshold
moving to a density below $n_{\rm sat}/2$.  The~high-density limit shows
the following new features: (a) the $\Lambda$ becomes the most
abundant baryon by~exceeding the neutron fraction; the $\Xi^0$
hyperon overtakes $\Xi^-$ and becomes the second-most abundant
hyperon. Note that the upper right panel of
Figure~\ref{fig:abundances-hyperons} differs from the panels shown here
only by the temperature ($T=50$~MeV); therefore, our comments here
parallel the statements made earlier in the context of
Figure~\ref{fig:abundances-hyperons}.  Turning to the $\Sigma$s, we note
that their abundances are noticeable for $T\geq 20$~MeV and the
occurrence of the special interchange point of isospin degeneracy is
seen again for $T=30$~MeV and $T=40$~MeV.

\begin{figure}[H] 
\includegraphics[width=.9\linewidth]{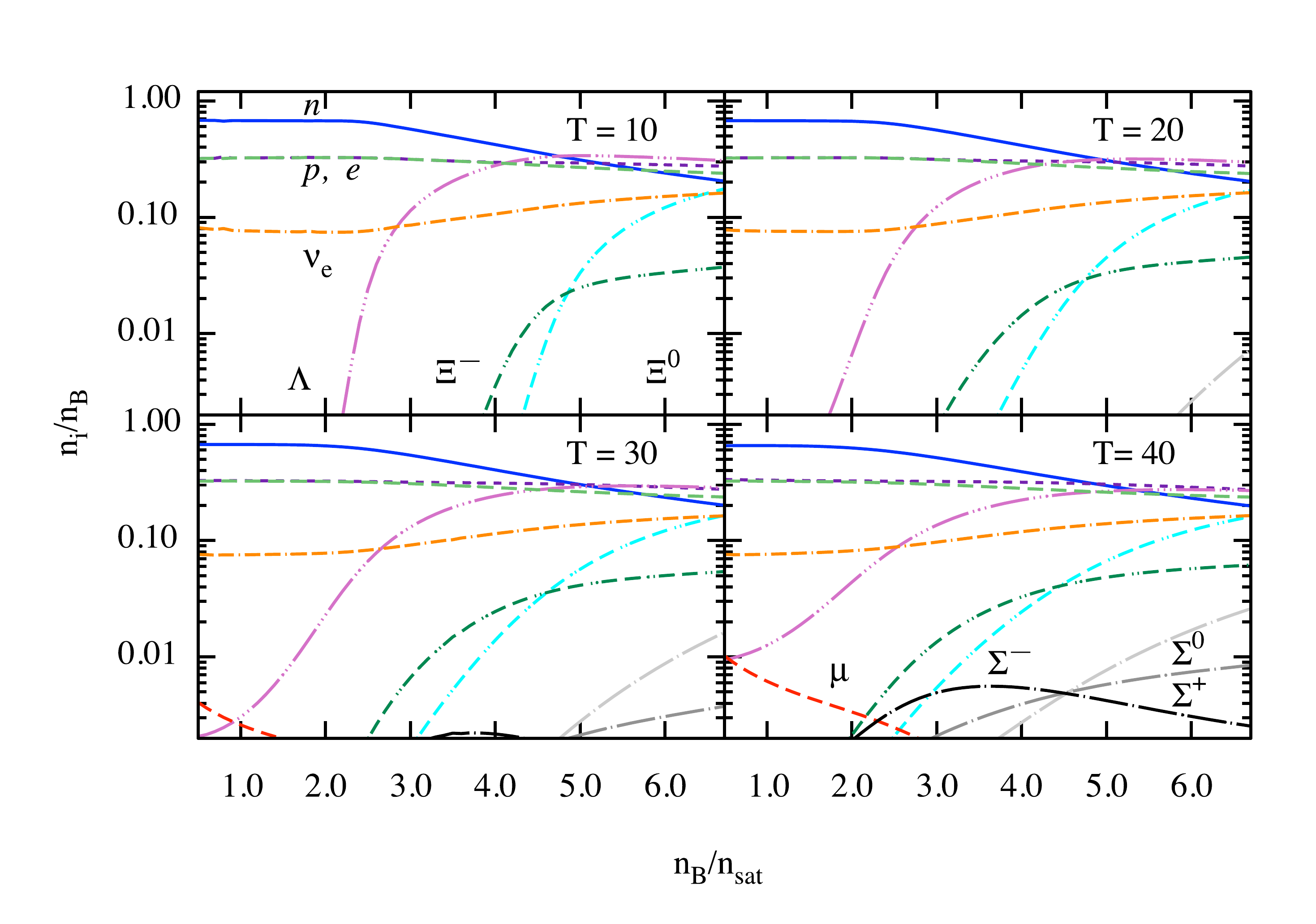}
\caption{ Same as in Figure~\ref{fig:abundances-hyperons-T_dep_Y_01},
  but for the fixed $Y_{L,e}= 0.4$ and $Y_{L,\mu}=0$, i.e.,~the lepton
  number fractions are characteristic for supernova.  
}
\label{fig:abundances-hyperons-T_dep_Y_04} 
\end{figure}

In Figure~\ref{fig:abundances-hyperons-T_dep_Y_04}, we show the same as
in Figure~\ref{fig:abundances-hyperons-T_dep_Y_01} but for
$Y_{L,e}= 0.4$ and $Y_{L,\mu}=0$, which physically corresponds to the
case of supernova matter. Many general trends seen for baryon
abundances remain the same under these new conditions. An~interesting
new feature is the near equipartition between neutrons, $\Lambda$,
 and protons at high density $n_B/n_{\rm sat} \ge 5$, with~
 $\Xi^0$ fraction approaching this group above $6n_{\rm sat}$. 
As for leptons,
the main effect arises from the drop of $\mu$-on fraction to below 1\%
and less for  $T\le 40$~MeV. For~$T=50$, this number climbs to a few
percent (see Figure~\ref{fig:abundances-hyperons}, lower
panels). Because~of this, the~charge neutrality is mainly maintained
by the equality of the abundances of protons and electrons,
with~slight disparity introduced by $\Xi^-$ at high density. The~most
striking difference is the strong enhancement of electron-neutrino
abundances for all temperatures, with~a very weak dependence on the
temperature of the~environment. 

\unskip

\section{Conclusions}
\label{sec:conclusions}

In this work, we explored the finite-temperature EoS of nuclear and
hypernuclear matter within the CDF formalism. Formally, our study uses
essentially the same approach as that of Ref.~\cite{Colucci2013}, but
it includes additional hidden-strangeness mesons and employs a
different strategy to fix the hyperonic couplings in the scalar sector
by adjusting these to the depths of hyperon potential in nuclear
matter. We performed parameter studies varying the temperature,
density and lepton fraction within two scenarios: the binary merger
remnant scenario with equal numbers of electron and $\mu$-on lepton
numbers and the supernova scenario with non-zero electron and zero
$\mu$-on lepton numbers. In~all cases, the~well-known feature of
softening of the EoS with the inclusion of hyperons is reproduced.
Even though the temperature dependence of the EoS is not strong (see
Figure~\ref{fig:EoS}), it has significant impact on the radii and
masses of compact stars (see for
example,~\cite{Raduta:2020fdn,Khadkikar:2021yrj}). The~abundances of
particles in a baryon--lepton mixture in a merger remnant and a
supernova were explored within the CDF formalism. The~main features
are: (a) at finite temperatures, the sharp increase in hyperon
fractions at the thresholds is replaced by a gradual increase over a
density range allowing for a significant fraction of hyperons,
especially $\Lambda$s, at~sub-saturation densities, as shown in
Figures~\ref{fig:abundances-hyperons},
\ref{fig:abundances-hyperons-T_dep_Y_01},
and \ref{fig:abundances-hyperons-T_dep_Y_04}. (b)
At large densities $n_B/n_{\rm sat} \ge 5$, the~most abundant baryon
is $\Lambda$, as~in the strongly relativistic regime, the difference
between the (bare) masses of the neutron and $\Lambda$ is not
important.  The~weaker coupling of $\sigma$ meson to $\Lambda$ than to
nucleon results in a a weaker renormalization of $\Lambda$ mass (see
Figure~\ref{fig:meff-hyperons}) which disfavors $\Lambda$
hyperons. However, the~weaker repulsive coupling of $\Lambda$s to
$\omega$-meson promotes their abundances compared to neutrons, which
eventually leads to their dominance at high densities. Note that the
$\rho$-meson coupling is exponentially suppressed at high densities
and it does not play any considerable role. Note also that the roles
played by $\sigma^*$- and $\phi$-mesons are similar to that of
$\sigma$- and $\omega$-mesons, but are quantitatively less
important. (c) The triplet of $\Sigma$ hyperons, which is completely
suppressed in the cold regime of hypernuclear matter, emerges at
temperatures above 20 MeV, with~significant fractions of $\Sigma^{-}$
compatible to that of $\Xi^{-}$ at low densities
$n_B\leq 2n_{\rm sat}$ and high temperatures $T\geq 40$~MeV.  (d) In
the neutrino-trapped regime, there is always a special isospin
degeneracy point where the charge chemical potential of the system
vanishes. At~that point, the~baryon abundances within each of the
three isospin-multiplets are equal to each other as a result of
(approximate) isospin symmetry.  (e) We find a significant difference
between the neutrino abundances in the merger remnant and supernova
cases. In~the first case, there are comparable numbers $\sim 1\%$ of
electron and $\mu$-on neutrinos (the electron and $\mu$-on lepton
numbers being equal). In~the second case, electron neutrino abundance
is much larger $\sim 10\%$ and $\mu$-on neutrinos are absent (there is
only a small fraction of $\mu$-on anti-neutrinos in this case,
typically less than a percent).  This, of~course, reflects the choices
of $Y_{L,e}$ and $Y_{L,\mu}$ for these cases, but~the abundances are
not trivially related to \mbox{lepton numbers.}

\vspace{6pt}

\authorcontributions{ A.S. and A.H. equally contributed
  to all stages of this project. All authors have read and agreed to the published version of the manuscript.
}

\funding{ This research was funded by the Volkswagen Foundation
  (Hannover, Germany) grant \mbox{No. 96 839}. A. S. was funded by Deutsche Forschungsgemeinschaft
(DFG) Grant No. SE1836/5-1. 
}
\institutionalreview{Not applicable. 
}

\informedconsent{Not applicable. 
}

\dataavailability{The data underlying this article will be shared on reasonable request to the corresponding author. 
}

\acknowledgments{ The author acknowledge the networking opportunities
  offered by the European COST Action “PHAROS” (CA16214).}

\conflictsofinterest{The authors declare no conflict of interest. 
} 



\end{paracol}

\reftitle{References}

\end{document}